# Ultrafast Terahertz Probe of Transient Evolution of Charged and Neutral Phase of Photoexcited Electron-hole Gas in Monolayer Semiconductor

Xuefeng Liu[1,2], Qingqing Ji[3], Zhihan Gao[1,2], Shaofeng Ge[1,2], Jun Qiu[1,2], Zhongfan Liu[3], Yanfeng Zhang[3,4], Dong Sun[1,2,*]

[1]International Center for Quantum Materials, School of Physics, Peking University, Beijing 100871, P. R. China
[2]Collaborative Innovation Center of Quantum Matter, Beijing 100871, P. R. China
[3]Center for Nanochemistry(CNC), Beijing National Laboratory for Molecular Sciences, College of Chemistry and Molecular Engineering, Academy for Advanced Interdisciplinary Studies, Peking University, Beijing 100871, P. R. China
[4]Department of Materials Science and Engineering, College of Engineering, Peking University, Beijing 100871, People's Republic of China
[*]Correspondance and request for materials should be addressed to D. S. (email: sundong@pku.edu.cn)

Abstract

We investigate the dynamical formation of excitons from photoexcited electron-hole plasma and its subsequent decay dynamics in monolayer $MoS_2$ grown by chemical vapor deposition using ultrafast pump and terahertz probe spectroscopy. Different photoexcited electron-hole states are resolved based on their distinct responses to THz photon and decay lifetime. The observed transient THz transmission can be fit with two decay components: a fast component with decay lifetime of 20 ps, which is attributed to exciton life time including the exciton formation and subsequent intraexciton relaxation; a slow component with extremely long decay lifetime of several ns due to either localized exciton state or a long live dark exciton state which is uncovered for the first time. The relaxation dynamics is further verified by temperature and pump fluence dependent studies of the decay time constants.

Monolayer transition-metal dichalcogenides (TMDC), as a new category of two dimensional(2D) materials, draw intense research interest in the post-graphene era due to their exceptional optoelectronic properties as 2D semiconductor counterpart of graphene[1-5] and versatile capability of quantum control of the spin and valley pseudospin through the Berry phase related properties and strong spin-orbit coupling[6-11]. Of central role governing these unique properties are the behaviors of the charge carriers in 2D TMDC, which are subjected to substantial coulomb interactions due to the strong quantum confinement and reduced screening in strict 2D limit. This leads the photoexcited electron-hole pairs to form an electron-hole bound state, known as exciton, which dominates the optoelectronics response and serves as carrier of various quantum degrees of freedom in 2D TMDC[6-8, 10, 11]. Recent experiment shows the tightly bound exciton can further capture additional excess charges to form trion (charged exciton)[12, 13]. Comparing to conventional bulk semiconductors and its low dimensional structure such as semiconductor quantum well[14], 2D TMDC processes very large binding energy up to a few hundred meV for exciton[15-20] and high dissociation energies up to 50 meV for trion[7, 12, 13, 21]. These values are an order of magnitude larger than those of their multilayer and bulk crystals[22]. Due to the extremely large binding energy, the evolution dynamics of neutral and charged phase from optically excited eletron-hole plasma is an interesting topic remaining to be elusive in TMDC, although it has been well studied in traditional semiconductors[23, 24].

In this letter, we apply ultrafast pump and terahertz probe spectroscopy on chemical vapor deposition grown monolayer $MoS_2$ sample to study the exciton formation and evolution dynamics from photo excited electron hole plasma. The schematic diagram of the experiment is shown in Fig. 1a: 3.1 eV (400 nm) or 1.55 eV (800 nm) pump photons are used to excite the sample through direct one-photon or two-photon interband optical transition. The photon energy is sufficient to produce electron hole plasma with above bandgap excitation after intervalley relaxation to the K (K') valley. After the excitation, a terahertz pulse coming at various delay time t probes the evolution of electron hole plasma by monitoring the pump induced terahertz transmission change through the sample. The response of THz photon to bound and unbound electron hole state is shown schematically in Fig. 1d. THz response of free carriers (Process I), whether from dopants or photoexcitation, can be essentially understood as the charge carriers driven by the alternating electric field of THz, this coupling can often be described with a Drude response function, or alternatively by modified Drude model in some special cases[23]. However, once electron and hole bind together to form exciton state, its coupling with terahertz field decreases substantially (Process II). This is because the exciton is neutral, so the terahertz field only couples weakly with exciton through resonant interactions with internal exciton transitions (intraexciton transition)[24] and through non-resonant interaction: the polarizability associated with the electron and hole wave functions of the exciton. Due to the large binding energy of exciton in $MoS_2$, the intraexciton transition energy from exciton ground state (1s) to first excited state (2p) is far larger than terahertz probe photon(<7 meV). The resonant intraexciton transition can happen between highly excited exciton states, whose occupations are very limited and only occurs during transient intraexciton relaxation. On the other hand, for non-resonant interaction, the exciton polarizaiton is also relatively small due to the short distance between the electron and hole in a tightly bound exciton. Additionally, this coupling can be further reduced if the exciton is localized by trapping center or evolves to state with even larger binding energy, such as a midgap dark state, as it is equivalent to effective mass increase (Process III).

To perform terahertz time domain pump probe spectroscopy measurement, a 250 kHz Ti-sapphire amplifier (RegA) system[25] is employed to generate laser pulses with 800 nm (1.55 eV) and 60 fs. The laser was split into three beams: one beam is either directly used or frequency doubled with a BBO crystal for ultrafast pump; the second beam is used to generate THz through a GaAs photoconductive switch; the third beam is used to mapping out the THz electric field waveform in the time domain through a 1 mm thick ZnTe crystal using standard electro-optic sampling technique[26]. The effective bandwidth of the sampling system is limited to 1.7 THz by phasematching in the ZnTe crystal and absorption of fused silica windows of cryostat. In our measurement, the generated THz combined with the optical pump beam are overlapped on the sample with a 2.0-mm and 1.5-mm (FWHM intensity) spot size respectively.

The large-area monolayer $MoS_2$ samples are grown by chemical vapor deposition (CVD) on sapphire substrate[27]. The fabrication detail of these CVD samples is provided in reference[27]. The left inset of Fig. 1(c) shows the color contrast of sapphire substrate with and without monolayer MoS$_2$ on it. The right inset of Figure 1(b) shows TEM image of the sample transferred on copper grids, wherein an unintentional scratch is utilized to identify the monolayer nature of the film. The CVD sample we used has over 96% monolayer coverage as verified from the PL mapping in Fig. 1c showing homogeneous intensity in the length scale of tens of microns. Electron doping density of a typical sample can be estimated from transport measurement, which is less than 7*10$^{11}$ cm$^{-2}$. This doping level is an order magnitude smaller than the samples used in a recent THz probe experiment on CVD grown MoS$_2$[28]. For comparison, the same THz probe experiment is also performed on thick MoS$_2$ samples (>100 nm) exfoliated from natural minerals and transferred on the sapphire substrate.

Figure 2a shows typical terahertz field waveform through a bare substrate, monolayer MoS$_2$ and bulk MoS$_2$ on sapphire. The change of THz peak field induced by monolayer MoS$_2$ is about 2.3%. In the time resolved measurement, we fix the THz sampling delay at the THz peak field and scan the delay time t between the pump and THz pulse. After 400 nm pump pulse excitation, we observe about 0.124% THz transmission decrease for monolayer and 4.19% for bulk (Fig. 2b), indicating a transient increase of THz absorption due to the photoexcited electron hole plasma. The rise time of the response is about 2 ps in monolayer (Fig. 2d), which is attributed to the scattering from initial excited C band to K (K') valley, The rising time is reduced to 1.5 ps in bulk due to different band structure. Fig. 2c shows the temporal evolution of the fractional change of THz field ΔE(t,0)/ E(0), where ΔE(t,0) is pump induced transmitted THz peak field change at delay t and E(0) is the transmitted THz field peak field with no pump excitation. The pump fluence of 400 nm and 800 nm are both 10 μJ/cm$^2$, comparing to 0.124% of ΔE(t,0)/ E(0) with one photon excitation, the ΔE(t,0)/ E(0) signal is reduced to 0.086% with two photon excitation. In all cases, the THz dynamics can be fit with biexponential decay with two decay components, a fast component $\tau_1$ and a slow component $\tau_2$. Both components are faster in bulk than those in monolayer. The temperature dependence of $\tau_1$ and $\tau_2$ are shown in Fig. 3, while $\tau_2$ increases as temperature increases, $\tau_1$ is relatively inert with temperature change in both samples. For pump fluence dependent measurement (Fig. 4), both $\tau_1$ and $\tau_2$ decreases as the pump fluence increases.

Now we turn to the interpretation of the transient THz transmission signal and its related decay dynamics observed in the experiment. The initial negative THz transmission signal around time zero can be explained by the free carrier absorption of THz due to the photoexcitaion of electron hole plasma by the pump pulse, as marked by process I in Fig. 1b. However, an attempt to fit the conductivity of monolayer with pure Drude model fails, possibly due to the low doping intensity and intial intervalley scattering process as marked by process II in Fig. 1b. The one photon absorption rate of 400 nm is measured to be around 30% by comparing the power transmission through MoS$_2$/sapphire and sapphire (Fig. 2e). This extremely large monolayer absorption is due to a band nesting effect in MoS$_2$ which has been discussed extensively in recent literature[29, 30]. With 0.32 μJ excitation energy, the absorption rate converts to photo excited electron hole density of 4.4*10$^{12}$/cm$^2$. For two photon absorption of 800 nm, the measured absorption coefficient is 2.3%, which converts to electron hole density of 5*10$^{11}$/cm$^2$. The amplitude of transient THz signal at timezero doesn't increase monotonically with excitation density when switching the pump wavelength from 400 nm to 800 nm. This is possibly due to many body effects during the initial stage, which causes different response under two-photon and one-photon excitation.

After the photoexcitation, the excited hot electron-hole plasma relaxes and starts to form its bound state due to the strong coulomb interaction (process III in Fig. 1b). As the electrons and holes bind together and form exciton, their responses to THz decreases and the transient THz signal starts to recover from the negative minimum. We attribute $\tau_1$ to be the exciton lifetime starting from its initial formation from electron hole plasma to its stay in Rydberg series, including relaxation process from excited state to ground state and subsequent relaxation to mid-gap state or to recombine through radiative or nonradiative process[31, 32]. Among all these processes, we infer the exciton life time is the slowest and dominates $\tau_1$. Recent measurement on monolayer $MoS_2$ on BN substrate gives 50 ps exciton lifetime. Considering $MoS_2$ on BN have better mobility and less defects, the measured 20 ps lifetime with THz probe are within reasonable range. Additionally, $\tau_1$ decreases as pump fluence increases, indicating exciton-exciton annihilation process[33] under the experimental excitation conditions. Exciton formation and intraexciton relxation time should be significant shorter than $\tau_1$. The exciton formation time has been measured to be within 2 ps in $MoS_2$[34], which is significantly faster than that of GaAs quantum well[24], possibly due to its extremely large binding energy. In both monolayer and bulk $MoS_2$, $\tau_1$ shows very weak lattice temperature dependence (Fig. 3). This is because the intraexciton relaxation, which is phonon scattering related, is estimated to be 4 ps in monolayer as measured on similar sample[34]. Another process could also contribute to the transient THz recovery during the fast decay process is the formation of a trion state which is claimed to be within ps[28] A trion state dulls the response to THz due to its increased effective mass compared to free carriers, thus increase the THz transmission. This effect has been discussed extensively in a recent transient THz measurement on similar sample but with an order of magnitude larger doping intensity than the monolayer sample used here. With initial high doping intensity, exciton favors to form trion compared to our sample, which may account for the opposite transient THz sign and totally different decay dynamics observed in their work[28].

The decay of transient THz response slows down significantly and follows another exponential decay with lifetime $\tau_2$ on the order of nanosecond in monolayer. According to the extremely long lifetime, temperature and pump fluence dependence, we attribute $\tau_2$ to be life time of exciton captured by defect states or relaxing to dark exciton state in the midgap as marked by process III in Fig.1b. In either case, the exciton state becomes highly localized, which further reduces the exciton's coupling to THz. The trapped exciton emission lifetime has been measured to be 125 ps at 4K by time resolved photoluminescence experiment[32], $\tau_2$ is an order of magnitude larger possibly due to higher temperature ($\tau_2$ increases as temperature increases), lower excitation density ($\tau_2$ decreases as pump fluence increases) and longer nonradiative lifetime (compare to radiative lifetime in TRPL measurement) in our measurement. As shown in Fig. 1d, the trapping of exciton by the defect state and subsequent non-radiative decay of the trapped exciton state should involve phonon emission process. As the phonon occupation increases with temperature, the nonradiative decay process is quenched. This explains the measured temperature dependence of $\tau_2$. The pump fluence dependence is mainly due to the limited number of trapping state, which limits the available trapped exiton state. Excitons that are not trapped by the trapping center has shorter decay lifetime which reduces the measured $\tau_2$. Another possible decay path of exciton is a long lived dark exciton state (Fig. 1b), which could correspond to forbidden exciton transition. This dark state has recently been theoretically considered[35] and experimentally studied by polarization resolved spectrum hole burning experiment[36]. The dark exciton lifetime is measured to be 6.3 ns in monolayer $MoSe_2$ and decreases as excitation intensity increases which also match our pump fluence dependent measurement.

In summary, we have studied the exciton formation and decay dynamics from photoexcited electron hole plasma through transient THz measurement. Different bound and unbound phases of photoexcited electron hole state that is sensitive to THz field can be resolved clearly in our measurement with two decay components in monolayer $MoS_2$: a fast component (~20ps) attributed to exciton formaton and life time; a long component(~ ns) attributed to intermediate midgap trapped or dark exciton state. We expect similar dynamics applies to other monolayer TMDC due to the similarity in bandstructure and exciton properties. The measured time resolved evolution of photoexcited carriers enriches our understanding of basic optoelectronic properties of 2D TMDC and provides new opportunities in developing novel optoelectronic and excitonic devices based on 2D TMDC.

**Acknowledgement:**
The authors want to acknowledge Xiaodong Xu for helpful discussion. This project has been supported by the National Basic Research Program of China (973 Grant Nos. 2012CB921300,2014CB920900, 2013CB932603, 2011CB921903, 2012CB921404, 2012CB933404 and 2011CB933003), the National Natural Science Foundation of China (NSFC Grant Nos. 11274015, 51222201, 51290272, 11304053, 51121091), the Recruitment Program of Global Experts, the Specialized Research Fund for the Doctoral Program of Higher Education of China (Grant No.20120001110066) and Beijing Natural Science Foundation (Grant No. 4142024).

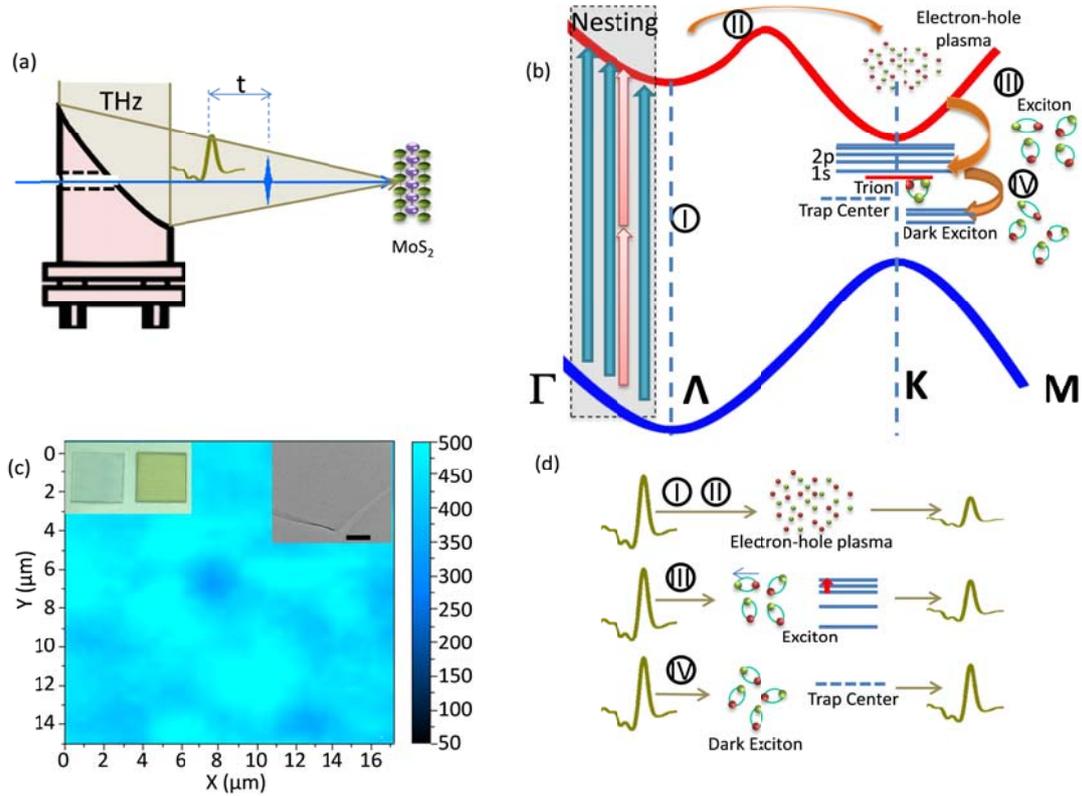

**Figure 1. Ultrafast pump and terahertz probe spectroscopy on CVD grown monolayer MoS$_2$**
**a**, Experimental setup. Terahertz radiation and pump pulse is focused and overlapped on the MoS$_2$ sample by a parabolic mirror with a small hole at the center. **b,** Relaxation dynamics of electron hole plasma after one-photon or two-photon photoexcitation. **c,** PL mapping at 670 nm of MoS$_2$ sample with 532-nm CW excitation. The left insert is optical image contrast of a bare sapphire substrate(left) and a sapphire substrate covered by monolayer MoS$_2$ (right). The sample size is 10mm×10mm×0.5mm. The right insert is TEM image of MoS$_2$ sample. The scale bar is 200 nm. **d,** THz response of electron-hole plasma, exciton and dark exciton/trapping center: (I,II) increased terahertz absorption due to electron hole plasma; (III) reduced THz absorption due to formation of exciton; (IV) further reduced THz absorption due to dark exciton or trapped exciton state.

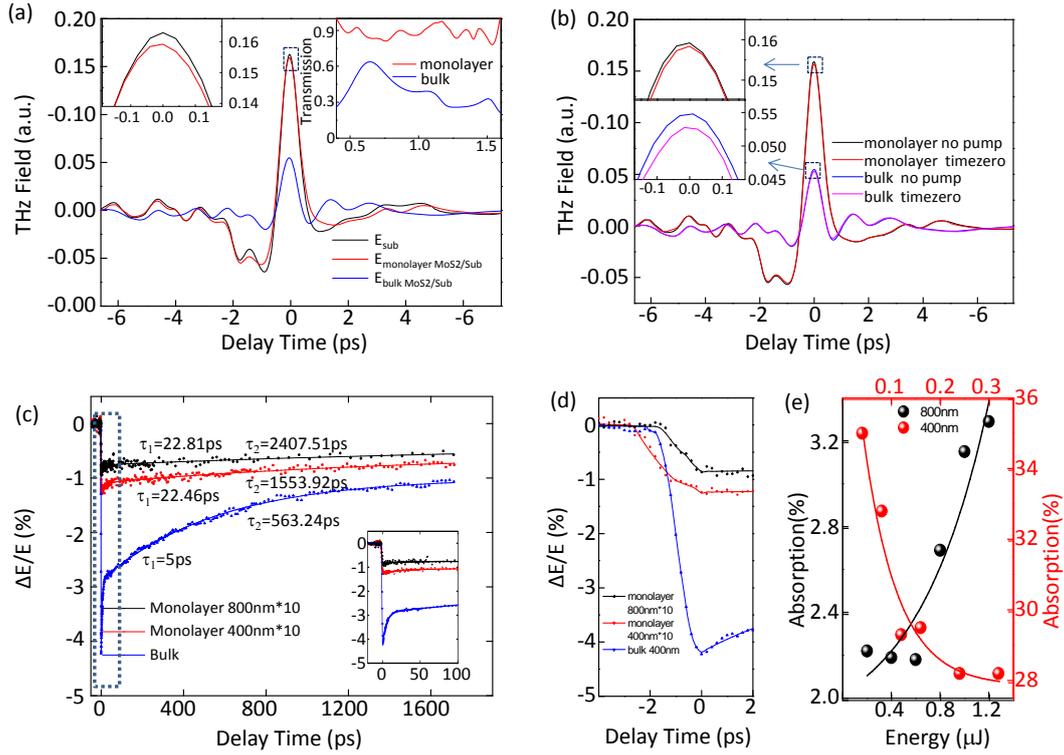

**Figure 2. Transient THz response of MoS$_2$ a,** THz transmission through monolayer, bulk MoS$_2$ samples and bare substrate in equilibrium condition. The left inset is a zoomed-in view of the terahertz field peak marked by rectangular in the main figure. The right inset shows the transmission spectrum of monolayer and bulk MoS$_2$. **b,** Pump induced reduction of THz transmission of monolayer and bulk MoS$_2$ at time zero. The pump light is 400nm, and the excitation energy is 0.32 μJ. The two inserts are zoomed-in views of the peak area as marked by rectangular area in the main figure. **c,** Temporal terahertz dynamics of monolayer and bulk MoS$_2$ with 400 nm and 800 nm excitation, the pump excitation fluence are the same. Experimental data is shown by dot and the line is bi-exponential decay fit. The insert is a zoomed-in view marked by the dash rectangular area. **d,** Zoom in of the rising part of transient THz signal in Fig. c. **e,** Absorption coefficient of monolayer MoS$_2$ at different pulse energies.

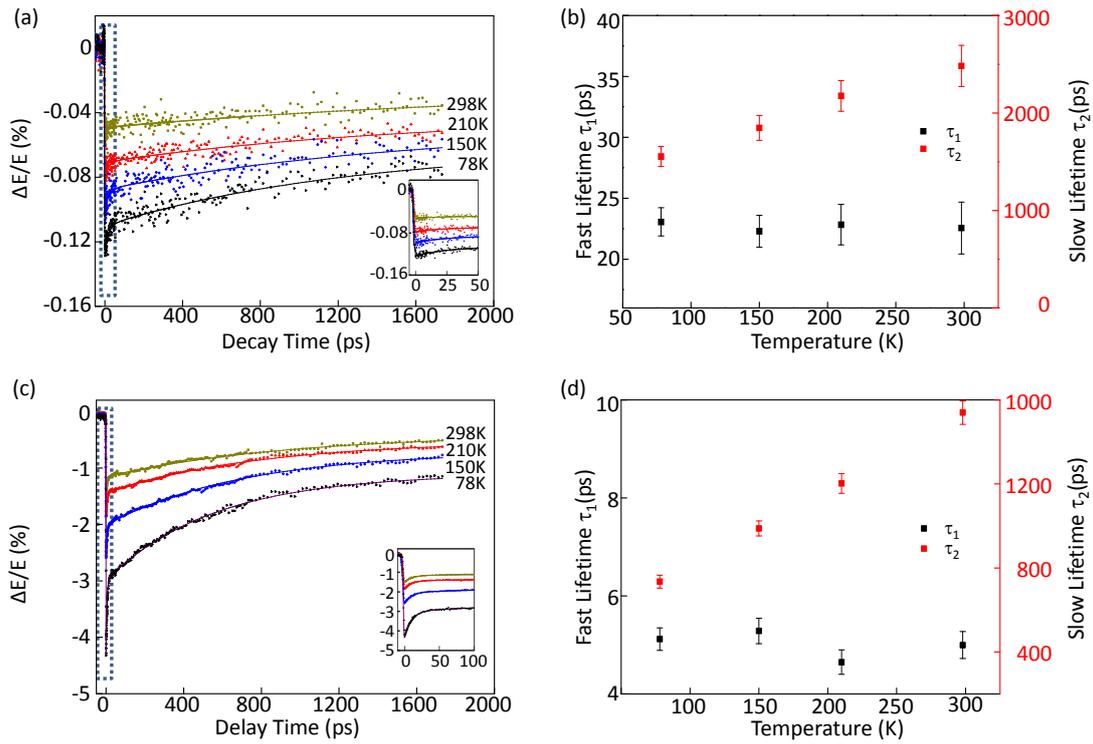

**Figure 3. Temperature dependence of terahertz response a,** Temporal evolution of Transient THz response of monolayer $MoS_2$ at 78K, 150K, 210K and 298K. The pump pulse is 400nm with 0.32 μJ. The experimental data is shown by dots and the lines are biexponential fit. The inset is a zoomed-in view of the fast component. **b,** Temperature dependence of $\tau_1$ and $\tau_2$ of monolayer $MoS_2$. **c,** Temporal evolution of transient THz response of bulk $MoS_2$ at 78K, 150K, 210K and 298K. **d,** Temperature dependence of $\tau_1$ and $\tau_2$ of bulk $MoS_2$.

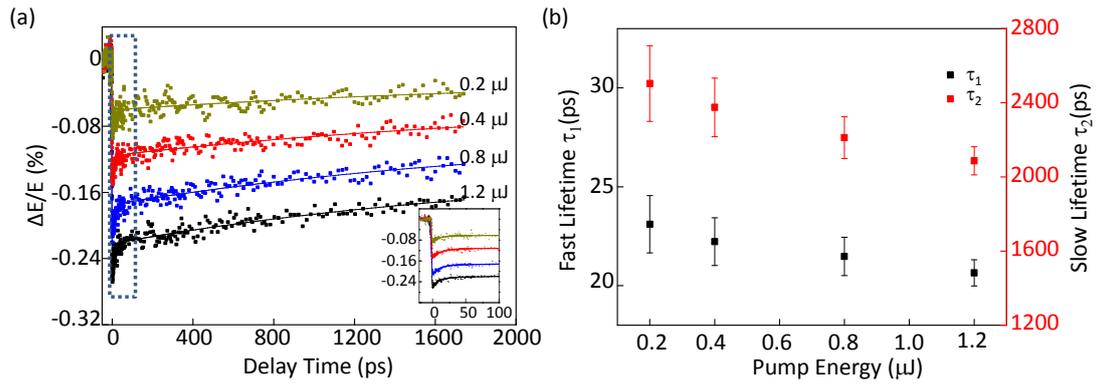

**Figure 4. Pump fluence dependence of terahertz response a,** Temporal evolution of transient THz of monolayer MoS$_2$ with 0.2 μJ, 0.4 μJ, 0.8 μJ and 1.2 μJ 800-nm pump pulse focused on 2-mm diameter spot. The inset is a zoomed-in view of the fast decay component. **b,** Pump pulse energy dependence of τ$_1$ and τ$_2$ of monolayer MoS$_2$.